\documentclass{article}
\usepackage{graphicx} 
\usepackage{caption}
\usepackage{subcaption}
\usepackage{wrapfig}
\usepackage{amssymb}
\usepackage{hyperref}
\usepackage{amsmath}
\usepackage{amsfonts}
\usepackage{mathtools}
\usepackage{bbm}
\usepackage{comment}
\usepackage{url}
\usepackage{mathtools}
\usepackage{amsthm}
\usepackage{float}
\usepackage{tikz}
\usepackage{multirow}
\usepackage{geometry}
\usetikzlibrary{spy}
\newcommand{\R}{\mathbb{R}}
\usepackage[preprint]{styleguide}
\usepackage{amsmath,graphicx}

\usepackage{enumitem}
\setlist{nosep, leftmargin=14pt}

\usepackage{mwe} 

\title{Author guidelines for ISBI proceedings manuscripts}
\name{Moritz Piening$^{1*}\quad$ Fabian Altekrüger$^{1,3}\quad$ Gabriele Steidl$^1\quad$Elke Hattingen$^2\quad$  Eike Steidl$^{2*}$\thanks{*Equal contributions
}
}
\address{$^1$Technische Universität Berlin$\quad^2$Goethe-Universität Frankfurt$\quad^3$Humboldt-Universität zu Berlin}

\title{Conditional Generative Models for Contrast-Enhanced Synthesis\\ of T1w and T1 Maps in Brain MRI}
\begin{document}
%
\maketitle
\begin{abstract}
Contrast enhancement by Gadolinium-based contrast agents (GBCAs) is a vital tool for tumor diagnosis in neuroradiology. 
Based on brain MRI scans of glioblastoma
before and after Gadolinium administration, we address enhancement prediction by neural networks with two 
new contributions. 
Firstly, we study the potential of
generative models, more precisely conditional diffusion and flow matching, for uncertainty quantification in virtual enhancement.
Secondly, we examine the performance of T1 scans from quantitive MRI versus T1-weighted scans. In contrast to T1-weighted scans, these scans have the advantage of a physically meaningful and thereby comparable voxel range. To compare network prediction performance of these two modalities with incompatible gray-value scales,
we propose to evaluate segmentations of contrast-enhanced regions of interest 
using Dice and Jaccard scores.
Across models, we observe better segmentations with T1 scans than with T1-weighted scans.
\end{abstract}


\section{Introduction}
\begin{figure}[t!]
\centering
\begin{subfigure}[t]{.195\linewidth}  
\begin{tikzpicture}[spy using outlines=
{rectangle,white,magnification=5,
width=1.7cm, height=1cm, connect spies}]
\node[anchor=south west,inner sep=0]  at (0,0) {\includegraphics[width=\linewidth]{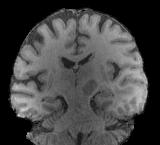}};
\spy on (1.00, .61) in node [right] at (-0.005,-.5);
\end{tikzpicture}
\end{subfigure}%
\hfill
\begin{subfigure}[t]{.195\linewidth}  
\begin{tikzpicture}[spy using outlines=
{rectangle,white,magnification=5,
width=1.7cm, height=1cm, connect spies}]
\node[anchor=south west,inner sep=0]  at (0,0) {\includegraphics[width=\linewidth]{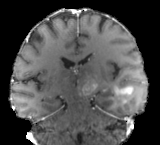}};
\spy on (1.00, .61) in node [right] at (-0.005,-.5);
\end{tikzpicture}
\end{subfigure}%
\hfill
\begin{subfigure}[t]{.195\linewidth}  
\begin{tikzpicture}[spy using outlines=
{rectangle,white,magnification=5,
width=1.7cm, height=1cm, connect spies}]
\node[anchor=south west,inner sep=0]  at (0,0) {\includegraphics[width=\linewidth]{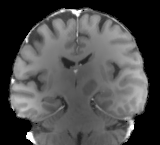}};
\spy on (1.00, .61) in node [right] at (-0.005,-.5);
\end{tikzpicture}
\end{subfigure}%
\hfill
\begin{subfigure}[t]{.195\linewidth}  
\begin{tikzpicture}[spy using outlines=
{rectangle,white,magnification=5,
width=1.7cm, height=1cm, connect spies}]
\node[anchor=south west,inner sep=0]  at (0,0) {\includegraphics[width=\linewidth]{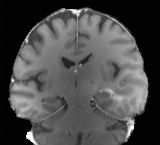}};
\spy on (1.00, .61) in node [right] at (-0.005,-.5);
\end{tikzpicture}
\end{subfigure}%
\hfill
\begin{subfigure}[t]{.195\linewidth}  
\begin{tikzpicture}[spy using outlines=
{rectangle,white,magnification=5,
width=1.7cm, height=1cm, connect spies}]
\node[anchor=south west,inner sep=0]  at (0,0) {\includegraphics[width=\linewidth]{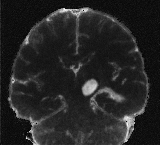}};
\spy on (1.00, .61) in node [right] at (-0.005,-.5);
\end{tikzpicture}
\end{subfigure}%
\vspace{.4mm}
\begin{subfigure}[t]{.195\linewidth}  
\begin{tikzpicture}[spy using outlines=
{rectangle,white,magnification=5,
width=1.7cm, height=1cm, connect spies}]
\node[anchor=south west,inner sep=0]  at (0,0) {\includegraphics[width=\linewidth]{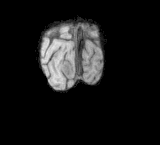}};
\spy on (.7, .8) in node [right] at (-0.005,-.5);
\end{tikzpicture}
\caption*{\scriptsize{Pre}}
\end{subfigure}%
\hfill
\begin{subfigure}[t]{.195\linewidth}  
\begin{tikzpicture}[spy using outlines=
{rectangle,white,magnification=5,
width=1.7cm, height=1cm, connect spies}]
\node[anchor=south west,inner sep=0]  at (0,0) {\includegraphics[width=\linewidth]{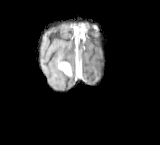}};
\spy on (.7, .8) in node [right] at (-0.005,-.5);
\end{tikzpicture}
\caption*{\scriptsize{Post}}
\end{subfigure}%
\hfill
\begin{subfigure}[t]{.195\linewidth}  
\begin{tikzpicture}[spy using outlines=
{rectangle,white,magnification=5,
width=1.7cm, height=1cm, connect spies}]
\node[anchor=south west,inner sep=0]  at (0,0) {\includegraphics[width=\linewidth]{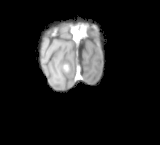}};
\spy on (.7, .8) in node [right] at (-0.005,-.5);
\end{tikzpicture}
\caption*{\scriptsize E2E}
\end{subfigure}%
\hfill
\hfill
\begin{subfigure}[t]{.195\linewidth}  
\begin{tikzpicture}[spy using outlines=
{rectangle,white,magnification=5,
width=1.7cm, height=1cm, connect spies}]
\node[anchor=south west,inner sep=0]  at (0,0) {\includegraphics[width=\linewidth]{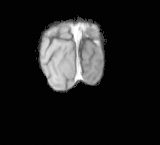}};
\spy on (.7, .8) in node [right] at (-0.005,-.5);
\end{tikzpicture}
\caption*{\scriptsize FM Mean}
\end{subfigure}%
\hfill
\begin{subfigure}[t]{.195\linewidth}  
\begin{tikzpicture}[spy using outlines=
{rectangle,white,magnification=5,
width=1.7cm, height=1cm, connect spies}]
\node[anchor=south west,inner sep=0]  at (0,0) {\includegraphics[width=\linewidth]{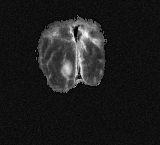}};
\spy on (.7, .8) in node [right] at (-0.005,-.5);
\end{tikzpicture}
\caption*{\scriptsize FM StdDev}
\end{subfigure}%
\caption{Two zoomed-in examples of HGG enhancement prediction with bad end-to-end (E2E) network performance.
E2E does not enhance the tumor region at all (first example) or only a small part (second example).
Flow matching (FM) mean shows blurred enhancement, but the standard deviation (StdDev) clearly indicates the uncertainty in the tumor area.
}
\label{fig_2}
\end{figure}

T1-weighted (T1w) MRI sequences after the administration of 
GBCAs play a crucial role in the diagnosis of brain tumors \cite{tsui2024reducing}. 
Yet, GBCAs are under discussion due to the possible retention in tissues and the additional acquisition scan time.
Therefore, it is an ongoing quest to eliminate Gadoli\-nium administration with synthetic MR predictions or to allow for lower doses by employing 
neural networks \cite{mallio2023artificial}.
Usual (end-to-end (E2E)) neural networks provide synthesized 
post-contrast images either from the pre-contrast scans only \cite{mallio2023artificial,preetha2021deep}, or using both  pre-contrast and  low-dose
scans \cite{haase2023artificial, pasumarthi2021generic, pinetz2024differential}.
The second method does not reduce the scan time 
but improves the prediction of small enhancement areas significantly. However, such areas play a smaller role in glioblastoma examination. For a comprehensive comparison of these two methods
including different MRI modalities (T1w, T2w, FLAIR) we refer to
\cite{bone2021contrast}.
\begin{figure}[t]
    \centering
    \includegraphics[width=\linewidth]{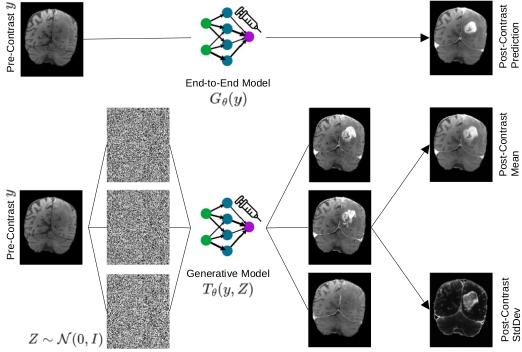}
    \caption{Illustration of E2E network (top) versus conditional generative one (bottom).
    While E2E networks provide only one contrast enhanced prediction for a pre-contrast image $y$, generative models 
    produce many samples (three depicted) from the distribution of post-contrast images conditioned to $y$. The sample mean gives a contrast prediction and their standard deviation shows areas of uncertainty. }
    \label{fig:overview}
\end{figure}
For this paper, we had access to MRI brain scans before and after Gadoli\-nium administration, where in addition to the T1w sequences, also T1 sequences from quantitative MRI (qMRI)
were available. T1 qMRI provides the actual T1 time and consequentially a meaningful voxel range, whereas T1w offers only relative differences without comparable values. Thus, T1 qMRI \emph{eradicates the need for voxel normalization} as necessary in T1w. Among others, this is of special interest for multi scanner or multi center studies \cite{noth2020quantitative}.
Nevertheless, E2E image synthesis may suffer from hallucinations and enhancement errors, see Fig. \ref{fig_2}, third column. 
Conditional generative neural networks produce  samples from the posterior distribution of enhanced images given the actual pre-enhanced scan.
Then their mean (average) can be used for enhancement
prediction, while their standard deviatiation quantifies uncertainty \cite{hagemannposterior}.
The basic idea is illustrated in Fig. \ref{fig:overview}.
Two state-of-the-art generative networks are diffusion \cite{song2021scorebased} and flow matching \cite{lipman2022flow} models.
\\
\textbf{Main contributions.}  We study for the first time
\begin{itemize}
    \item  conditional generative models, 
    more precisely diffusion models and flow matching, 
    for contrast-enhanced scan synthesis 
    and uncertainty quantification in brain MRI;
    \item applications of neural networks for contrast enhancement in T1-qMRI scans including a segmentation-based
    comparison to T1w-MRI by Dice and Jaccard scores. 
\end{itemize}
\section{Methods}
First, we outline the problem of virtual enhancement and the two MRI modalities in question. Next, we brefly describe the used E2E and the two generative neural network models.
\subsection{Contrast-Enhancement in T1w and T1}
Given pairs of pre-contrast and post-contrast 3D MR scans in $\R^{H\times W \times D}$, we follow \cite{pasumarthi2021generic} and extract 2.5D pre-contrast slices $y \in \R^{\tilde{d}} = \R^{H \times W \times 7}$ and 2D post-contrast slices $x \in \R^{d} = \R^{H\times W}$ with central slice of $y$ aligned to $x$.  
Considering 2.5D cubes consisting of multiple axial slices allows us to ensure spatial robustness and to account for the depth dimension.
As in \cite{pinetz2024differential}, we use the voxel-wise difference between pre- and post-contrast scans for learning neural networks
to predict $x$ given $y$.
We deal with two MRI modalities.
\\
\textbf{T1w MRI} produces images, where tissues with short T1 relaxation times (like fat) appear bright, while those with long T1 times (like water) appear darker. 
Despite excellent qualitative images, voxel-wise values do not carry physical meaning and depend on the scanner setup. Consequently, voxel normalization is an important pre-processing step. Here, we scale each pre- and post-contrast pair by dividing with the maximal pre-contrast 2.5D slice voxel.
\\
\textbf{T1 qMRI} scans, based on calculated relaxometry, measure the actual T1 relaxation times of tissues. This leads to a meaningful voxel range. T1 relaxation times are calculated by acquiring multiple MR images, 
fitting the signal intensities to an exponential recovery model, and deriving T1 values from this curve \cite{preibisch2009influence}. No rescaling is necessary.

Tissue voxels in T1w and T1 are inversely proportional, i.e.,
contrast-enhanced T1w regions become brighter, and contrast-enhanced T1 regions become darker.

\begin{table*}
    \centering
    \resizebox{\linewidth}{!}{
    \begin{tabular}{l||ccc|cc||ccc|cc}
    \textbf{Model/Data} &  \multicolumn{3}{c|}{HGG-T1w} & \multicolumn{2}{c||}{MET-T1w} &  \multicolumn{3}{c|}{HGG-T1} & \multicolumn{2}{c}{MET-T1} \\
         & MAE$\downarrow$ & rMAE$\downarrow$ & SSIM$\uparrow$ & MAE$\downarrow$ & SSIM$\uparrow$ & MAE$\downarrow$ & rMAE$\downarrow$ & SSIM$\uparrow$ & MAE$\downarrow$ & SSIM$\uparrow$\\
         \hline
    DM Mean& .027$\pm$.06&.074$\pm$.11&.813$\pm$.03 & .028$\pm$.06 & .829$\pm$.04 & .022$\pm$.05&.037$\pm$.05&.813$\pm$.03 & .026$\pm$.06 & .790$\pm$.05\\
    FM Mean&.026$\pm$.05&.066$\pm$.10&.756$\pm$.09&.027$\pm$.05&.754$\pm$.07&.021$\pm$.05&.035$\pm$.05&.797$\pm$.05& .025$\pm$.05&.791$\pm$.06\\
    E2E&.022$\pm$.05&.066$\pm$.10&.890$\pm$.03&.023$\pm$.05&.892$\pm$.04&.018$\pm$.05&.034$\pm$.05&.888$\pm$.03&.022$\pm$.05&.872$\pm$.04\\
    \hline
    Pre-Contrast&.037$\pm$.09&.095$\pm$.14&.827$\pm$.04&.038$\pm$.09&.828$\pm$.04&.033$\pm$.08&.053$\pm$.07&.814$\pm$.04&.041$\pm$.10&.792$\pm$.04\\
    \end{tabular}}
    \caption{Quantiative prediction evaluation of diffusion (DM), flow matching (FM) and end-to-end (E2E) models, where $\downarrow$ means `lower is better' and $\uparrow$ `higher is better'. On average E2E gives the best results, followed by DM and FM.}
    \label{tab:eval_rec}
\end{table*}
\subsection{Neural Networks} 
\textbf{End-to-End Models}  train a neural network $G_\theta \colon \R^{\tilde{d}} \to \R^d$ that approximates a mapping from pre-contrast scans to full-dose post-contrast scans based on empirical pairs 
$\{(x_i, y_i)\}_{i=1}^n \sim P_{X, Y}$,
where $P_{X,Y}$ denotes the joint distribution of random variables $X \in \R^d$ and $Y \in \R^{\tilde d}$.
Training can be achieved by minimizing a voxel-wise loss, here the mean absolute error (MAE)
\begin{equation*}
    \mathcal{L}(\theta) = \mathbb{E}_{(x,y) \sim P_{X,Y}} \big[ \Vert G_\theta (y) - x \Vert_1 \big].
\end{equation*}
For each pre-contrast scan $y$, such networks predict a single
post-contrast scan $x = G_\theta(y)$, see Fig. \ref{fig:overview}. For alternative loss function choices see \cite{pasumarthi2021generic}.
\\[1ex]
\textbf{Conditional Generative Models} learn, based on samples $\{(x_i, y_i)\}_{i=1}^n~\sim~P_{X, Y}$,
a conditional network $T_\theta(y, \cdot)$
that generates samples from the posterior distribution $P_{X|Y=y}$ 
by pushforwarding a latent Gaussian distribution $P_Z$, i.e.
$P_{X|Y=y} \approx T_\theta ^{-1}(y,\cdot)_\# P_Z \coloneqq P_Z (T_\theta ^{-1}(y,\cdot))$.
By sampling from our learned posterior distribution, we get several scan predictions, see Fig. \ref{fig:overview} bottom.
This allows insights into statistical properties like the mean and voxel-wise standard deviation, which provides a tool to quantify prediction uncertainty.
We examine two state-of-the-art generative models.
\\
1. \emph{Diffusion Models (DMs)}
\cite{song2021scorebased} employ stochastic differential equations, where the forward process transforms a target distribution to a Gaussian one by $\mathrm{d} X_t = - \frac{1}{2} \alpha_t X_t \mathrm{d} t + \sqrt{\alpha_t} \mathrm{d} W_t$, $X_0 \sim P_{X|Y =y}$,
where $W_t$ denotes the  Brownian motion and $\alpha_t$ is a positive, increasing noise schedule,
and the reverse process transforms
a Gaussian to the target distribution  by
\begin{align*}
     \mathrm{d}X_{T-t} 
     &= \alpha_{T-t}\big(\frac{1}{2} X_{T-t} +  \nabla \log p_{T-t} (X_{T-t}|Y=y) \big) \mathrm{d}t
     \\
   &\quad + \sqrt{\alpha_{T-t}} \mathrm{d}W_{T-t}, \quad X_T \sim \mathcal N(0,I_d).
\end{align*}
A neural network approximates the  conditional score $s_\theta \coloneqq \nabla \log p_t$
for a fixed noise levels $\alpha_t$
 by maximizing the `evidence lower bound' loss.
Given the score, we simulate the reverse process to transform Gaussian 
to posterior samples, see Fig. \ref{fig:overview} bottom.
Here, we use the image-to-image diffusion model \emph{Palette} \cite{palette}\footnote{unofficial code available at \url{https://github.com/Janspiry/Palette-Image-to-Image-Diffusion-Models}}.
\\
\emph{2. Flow Matching (FM)} \cite{chemseddine2024conditional,lipman2022flow} 
learns the vector field $v_t^\theta \colon \R^d \times \R^{\tilde{d}} \to \R^d$ of an 
ordinary differential equation for $t \in [0,1]$,
\begin{equation}\label{ode}
    \frac{\mathrm{d}}{\mathrm{d}t} \phi(x,y) = v_t (\phi_t (x,y)), \;
    \phi_0 (x,y) = (x,y),
\end{equation}
such that $\phi_t \colon \R^d \times \R^{\tilde{d}} \to \R^d$ defines a probability path $p_t = \phi_t (\cdot , y)_{\#}P_Z$, $t \in [0,1]$ and $P_{X|Y=y} = \phi_1 (\cdot,y)_{\#}P_Z$ based on the linear interpolation between $(Z,Y)$ and $(X,Y)$.
Once trained, we sample from $P_{X|Y=y}$ by solving \eqref{ode} with DOPRI-5.
We use our own FM implementation 
\footnote{available at Github after paper acceptance}.
\begin{table}[t!]
    \centering
    \resizebox{\linewidth}{!}{
    \begin{tabular}{l||cc|cc||cc|cc}
    \textbf{Model/Data}&  \multicolumn{2}{c|}{HGG-T1w} & \multicolumn{2}{c||}{MET-T1w} &  \multicolumn{2}{c|}{HGG-T1} & \multicolumn{2}{c}{MET-T1} \\
    Corr$\uparrow$ &AE&RE&AE &RE &AE &RE &AE &RE \\
    \hline
    DM StdDev&.643&.085&.637&.123&.680&.092&.647&.087\\
    FM StdDev&.631&.069&.636&.109&.668&.082&.637&.076\\
    \end{tabular}}
    \caption{Pearson correlation coefficient between voxel-wise standard deviation (StdDev) and the absolute (AE) and relative voxel errors (RE) for DM and FM means. Significance levels below 1\% are achieved for all correlation coefficients.}
    \label{tab:correlation}
\end{table}

\begin{figure}[h!]
\centering
\begin{subfigure}[t]{.04\linewidth}  
\centering
\rotatebox{90}{\scriptsize T1}
\end{subfigure}%
\hfill
\begin{subfigure}[t]{.135\linewidth}  
\includegraphics[width=\linewidth]{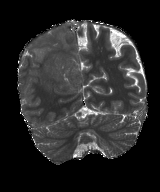}
\end{subfigure}%
\hfill
\begin{subfigure}[t]{.135\linewidth}  
\includegraphics[width=\linewidth]{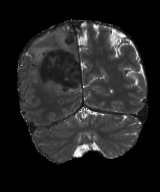}
\end{subfigure}%
\hfill
\begin{subfigure}[t]{.135\linewidth}  
\includegraphics[width=\linewidth]{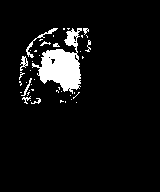}
\end{subfigure}%
\hfill
\begin{subfigure}[t]{.135\linewidth}  
\includegraphics[width=\linewidth]{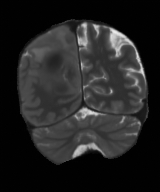}
\end{subfigure}%
\hfill
\begin{subfigure}[t]{.135\linewidth}  
\includegraphics[width=\linewidth]{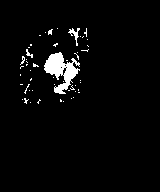}
\end{subfigure}%
\hfill
\begin{subfigure}[t]{.135\linewidth}  
\includegraphics[width=\linewidth]{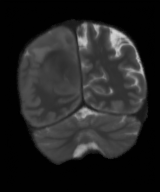}
\end{subfigure}%
\hfill
\begin{subfigure}[t]{.135\linewidth}  
\includegraphics[width=\linewidth]{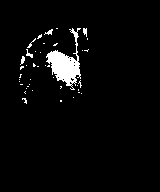}
\end{subfigure}%
\vspace{-.1mm}
\begin{subfigure}[t]{.04\linewidth}  
\centering
\rotatebox{90}{\scriptsize T1w}
\end{subfigure}%
\hfill
\begin{subfigure}[t]{.135\linewidth}  
\includegraphics[width=\linewidth]{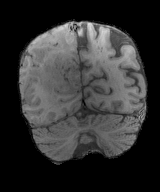}
\end{subfigure}%
\hfill
\begin{subfigure}[t]{.135\linewidth}  
\includegraphics[width=\linewidth]{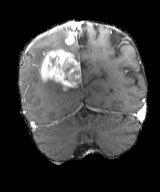}
\end{subfigure}%
\hfill
\begin{subfigure}[t]{.135\linewidth}  
\includegraphics[width=\linewidth]{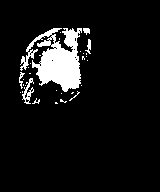}
\end{subfigure}%
\hfill
\begin{subfigure}[t]{.135\linewidth}  
\includegraphics[width=\linewidth]{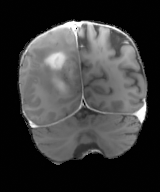}
\end{subfigure}%
\hfill
\begin{subfigure}[t]{.135\linewidth}  
\includegraphics[width=\linewidth]{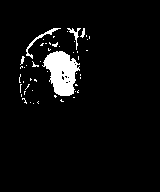}
\end{subfigure}%
\hfill
\begin{subfigure}[t]{.135\linewidth}  
\includegraphics[width=\linewidth]{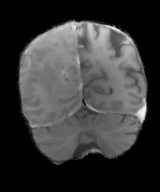}
\end{subfigure}%
\hfill
\begin{subfigure}[t]{.135\linewidth}  
\includegraphics[width=\linewidth]{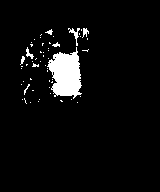}
\end{subfigure}%
\vspace{.4mm}
\begin{subfigure}[t]{.04\linewidth}  
\centering
\rotatebox{90}{\scriptsize T1w}
\end{subfigure}%
\hfill
\begin{subfigure}[t]{.135\linewidth}  
\includegraphics[width=\linewidth]{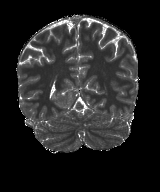}
\end{subfigure}%
\hfill
\begin{subfigure}[t]{.135\linewidth}  
\includegraphics[width=\linewidth]{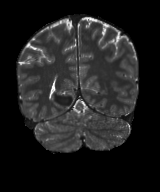}
\end{subfigure}%
\hfill
\begin{subfigure}[t]{.135\linewidth}  
\includegraphics[width=\linewidth]{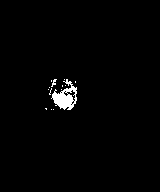}
\end{subfigure}%
\hfill
\begin{subfigure}[t]{.135\linewidth}  
\includegraphics[width=\linewidth]{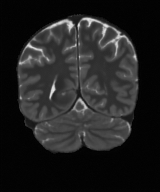}
\end{subfigure}%
\hfill
\begin{subfigure}[t]{.135\linewidth}  
\includegraphics[width=\linewidth]{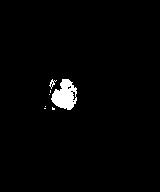}
\end{subfigure}%
\hfill
\begin{subfigure}[t]{.135\linewidth}  
\includegraphics[width=\linewidth]{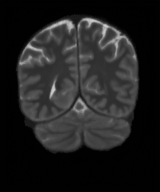}
\end{subfigure}%
\hfill
\begin{subfigure}[t]{.135\linewidth}  
\includegraphics[width=\linewidth]{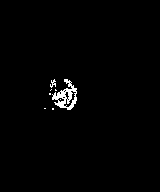}
\end{subfigure}%
\vspace{-.1mm}
\begin{subfigure}[t]{.04\linewidth}  
\centering
\rotatebox{90}{\scriptsize T1}
\end{subfigure}%
\begin{subfigure}[t]{.135\linewidth}  
\end{subfigure}%
\hfill
\begin{subfigure}[t]{.135\linewidth}  
\includegraphics[width=\linewidth]{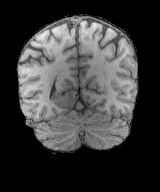}
\caption*{\scriptsize Pre}
\end{subfigure}%
\hfill
\begin{subfigure}[t]{.135\linewidth}  
\includegraphics[width=\linewidth]{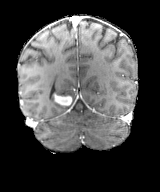}
\caption*{\scriptsize Post}
\end{subfigure}%
\hfill
\begin{subfigure}[t]{.135\linewidth}  
\includegraphics[width=\linewidth]{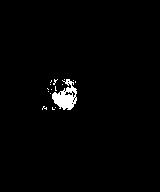}
\caption*{\scriptsize GT Seg}
\end{subfigure}%
\hfill
\begin{subfigure}[t]{.135\linewidth}  
\includegraphics[width=\linewidth]{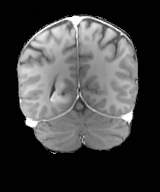}
\caption*{\scriptsize FM Mean}
\end{subfigure}%
\hfill
\begin{subfigure}[t]{.135\linewidth}  
\includegraphics[width=\linewidth]{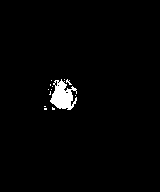}
\caption*{\scriptsize FM Seg}
\end{subfigure}%
\hfill
\begin{subfigure}[t]{.135\linewidth}  
\includegraphics[width=\linewidth]{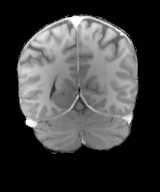}
\caption*{\scriptsize DM Mean}
\end{subfigure}%
\hfill
\begin{subfigure}[t]{.135\linewidth}  
\includegraphics[width=\linewidth]{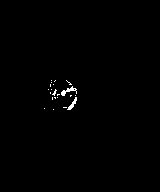}
\caption*{\scriptsize DM Seg}
\end{subfigure}%
\caption{Two examples of HGG enhancement prediction by FM, resp. DM and segmentation for T1 and T1w scans. 
Based on the difference between ground-truth pre- and post-contrast slices, the regions of interest and thresholding, we get comparable T1 and T1w ground-truth segmentations (3rd column, 20\% outlier threshold) which can be compared with  FM  (5th column) and  DM segmentations (7th column) via the Dice and Jaccard score.}
\label{fig:seg}
\label{tab:seg}
\end{figure}

\begin{figure}[h!]
    \centering
    \includegraphics[width=\linewidth]{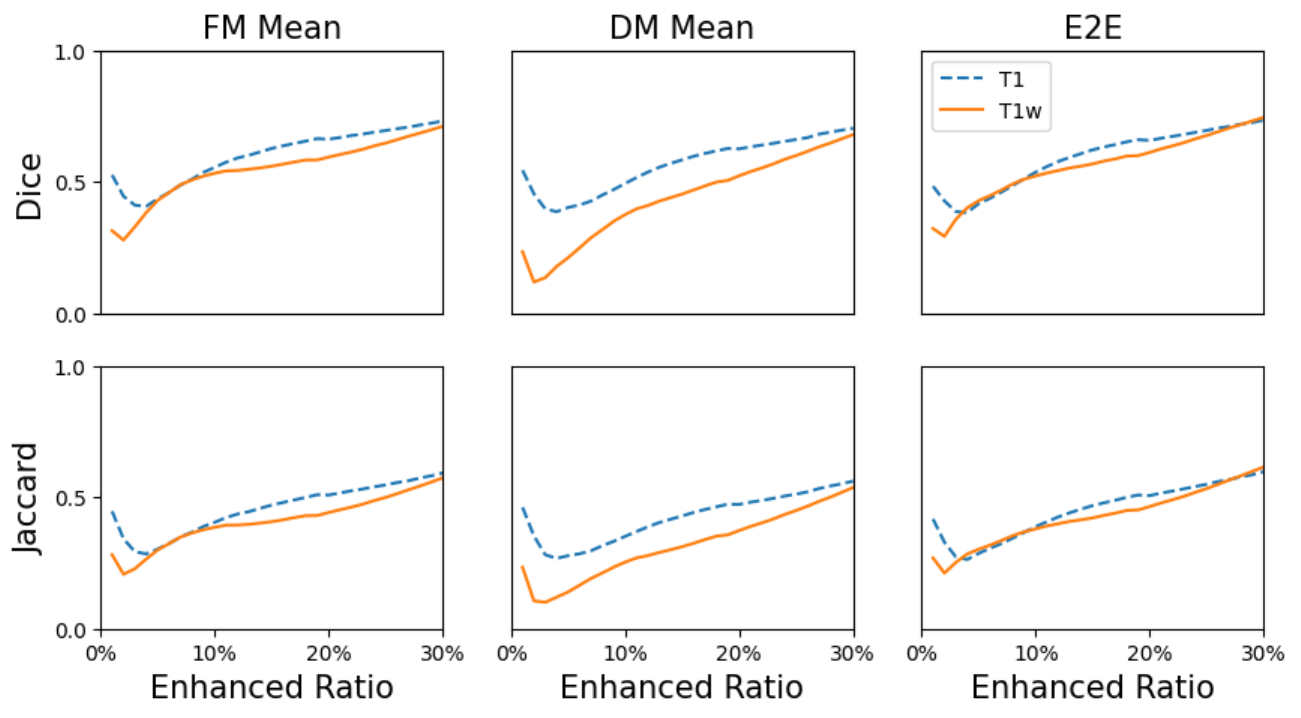}
    \caption{Segmentation comparison (Dice and Jaccard $\uparrow$) of contrast-enhanced region estimated by applying a voxel-wise difference outlier threshold between 0\% and 30\% to the ROI. T1 (blue dashed) gives better results than T1w (orange full).}
    \label{tab:segmentation}
\end{figure}
\section{Experiments}
We conduct the following experiments to analyze the benefit of conditional generative models and quantitative MRI.
\subsection{Dataset}
We used the MRI exams of 90 brain tumor patients acquired with a field strength of 3T (Magnetom Skyra-fit or Prisma, Siemens Healthineers, Erlangen, Germany); for the detailed MRI sequence parameters see \cite{noth2020quantitative}. 
Among these patients 69  have been diagnosed with high-grade gliomas (HGG) and 21  with metastases (MET).
For each exam, we have paired coregistered and skull-stripped pre-contrast and post-contrast T1w and T1 scans available.
 Focussing on HGG patients, we use a training set of axial slices from 64 HGG patients, where we remove peripheral slices without tumors. For each of the additional five HGG patients, segmentation masks with regions of interest (ROIs) were provided by a trained radiologist. For our HGG test set, we filter for slices that overlap with the ROIs. From the remaining slices, we use every fifth slice for the final HGG test (107 in total). 
Additionally, we extract four evenly-spaced central slices from each patient with metastases (MET). We use the slices of four randomly selected MET patients as a validation set. The 68 slices of the remaining 17 MET patients are a MET test set.
\subsection{Implementation}
We train each model using only T1, resp. T1w data
with the same U-Net architecture. We employ the network implementation of \cite{palette} with 32 channels, two residual blocks, 20\% dropout, and 16 attention head channels. 
The DM noise scales $\alpha_t$ are placed equidistantly between $1e^{-3}$ and $5e^{-2}$ for $2000$ steps. For FM, we provide our own implementation 
and discretize the ODE with 10 time steps and DOPRI-5. We train each model for 1500 epochs. We observe saturation, but no overfitting on the validation set due to extensive random rotation and cropping training augmentations.
For each 7-slice-stack $y$, we generate 50 image samples with DM and FM. 
\subsection{Evaluation Metrics}
We report the mean absolute error across all slices (MAE) and within the ROI (rMAE)  and the SSIM \cite{wang2004image} in Tab. \ref{tab:eval_rec}.
In Tab. \ref{tab:correlation}, we display the Pearson correlation coefficient between the estimated
FM StdDev and the absolute error $|\text{FM Mean}~-~\text{Post}|$ (AE), resp.
the relative error $\frac{|\text{FM Mean}~-~\text{Post}|}{|\text{Pre}-\text{Post}|}$ (RE). We skip the voxels where
$\text{Pre}-\text{Post} = 0$.
The same is computed for DM image samples.
To compare the network performance for T1w and T1 scans, which have incompatible gray value scales, we use thresholding in the ROI to segment contrast-enhanced regions. For each slice, we consider the voxel-wise absolute difference between the GT pre- and post-contrast slices and separate
a specific percentage (1\% - 30\%) of the voxels with the highest intensities, resulting in a binary segmentation, 
see Fig. \ref{fig:seg} third column. This is compared
with the same percentage of thres\-holded pixels
in the absolute difference between pre-contrast
and E2E, FM mean, DM mean, resp., see Fig. \ref{fig:seg} fifth and seventh column.
For the three cases, False Positives (FP), True Positives (TP), and False Negatives (FN) can be detected leading to 
the Dice $D$ and Jaccard $J$ scores   \cite{taha2015metrics} , 
$$
D = \tfrac{2 \text{TP}}{2 \text{TP} + \text{FP} + \text{FN}}, \quad
J = \tfrac{\text{TP}}{\text{TP} + \text{FP} + \text{FN}},$$
which range from 0 to 1 with 1 indicating perfect similarity. Note that the ground truth T1 and T1w segments should be identical for every threshold up to measurement errors due to the relation between T1 and T1w scans, see Fig. \ref{fig:seg}, third column. We choose thresholds of more than 30\% to exclude noise since only 42\% of ROI voxels show \textit{any} pre-post difference on average.
\subsection{Results}
In general, we observe a promising quality of all neural network-generated enhancement predictions, see Fig. \ref{fig:final}.
Tab. \ref{tab:eval_rec} shows \emph{on average}  (of 107 slices)
the best performance for E2E predictions, followed by DM and FM.
However, as shown in Fig. \ref{fig_2}, an E2E prediction may potentially be insufficient 
as often indicated by the standard deviation of a generative model output.
 Given the stronger pre-post differences in the last row of \ref{tab:eval_rec}, we see either no or only a slight reconstruction quality drop for the MET test set. 
 \\
 Moreover, we observe a high (Pearson) correlation between voxel errors and standard deviations
 in Tab. \ref{tab:correlation}. Consequently, errors are higher in regions of high standard deviation and vice versa. After accounting for the ground truth differences caused by contrast enhancement with the relative error, the correlation reduces.  This indicates that the correlation can partially be attributed to an increased uncertainty in regions of higher contrast enhancement. Nevertheless, the correlation coefficients remain positive indicating that the standard deviation also accounts for relative reconstruction errors.
 \\
As for the two MRI modalities T1w and T1, we see 
in Fig. \ref{tab:segmentation}
consistently better segmentation metrics for T1 slices across models, thresholds, and metrics. This may indicate an advantage of learning with T1-qMRI scans instead of T1w-MRI ones. Here FM gives better results than DM followed by E2E. 
\begin{figure}[h!]
\centering
\begin{subfigure}[t]{.04\linewidth}  
\centering
\rotatebox{90}{\scriptsize T1}
\end{subfigure}%
\hfill
\begin{subfigure}[t]{.137\linewidth}  
\includegraphics[width=\linewidth]{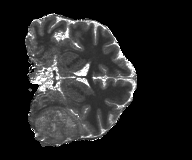}
\end{subfigure}%
\hfill
\begin{subfigure}[t]{.137\linewidth}  
\includegraphics[width=\linewidth]{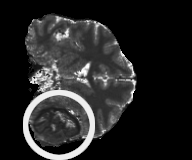}
\end{subfigure}%
\hfill
\begin{subfigure}[t]{.137\linewidth}  
\includegraphics[width=\linewidth]{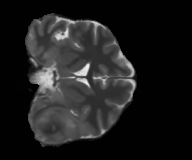}
\end{subfigure}%
\hfill
\begin{subfigure}[t]{.137\linewidth}  
\includegraphics[width=\linewidth]{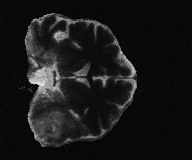}
\end{subfigure}%
\hfill
\begin{subfigure}[t]{.137\linewidth}  
\includegraphics[width=\linewidth]{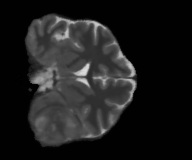}
\end{subfigure}%
\hfill
\begin{subfigure}[t]{.137\linewidth}  
\includegraphics[width=\linewidth]{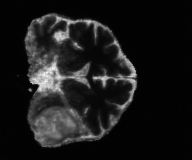}
\end{subfigure}%
\hfill
\begin{subfigure}[t]{.137\linewidth}  
\includegraphics[width=\linewidth]{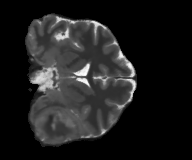}
\end{subfigure}%
\vspace{-.1mm}
\begin{subfigure}[t]{.04\linewidth}  
\centering
\rotatebox{90}{\scriptsize T1w}
\end{subfigure}%
\hfill
\begin{subfigure}[t]{.137\linewidth}  
\includegraphics[width=\linewidth]{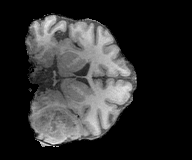}
\end{subfigure}%
\hfill
\begin{subfigure}[t]{.137\linewidth}  
\includegraphics[width=\linewidth]{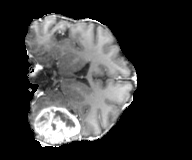}
\end{subfigure}%
\hfill
\begin{subfigure}[t]{.137\linewidth}  
\includegraphics[width=\linewidth]{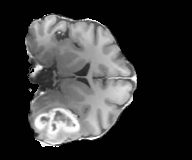}
\end{subfigure}%
\hfill
\begin{subfigure}[t]{.137\linewidth}  
\includegraphics[width=\linewidth]{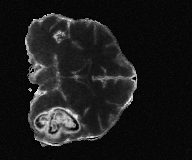}
\end{subfigure}%
\hfill
\begin{subfigure}[t]{.137\linewidth}  
\includegraphics[width=\linewidth]{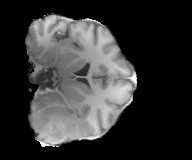}
\end{subfigure}%
\hfill
\begin{subfigure}[t]{.137\linewidth}  
\includegraphics[width=\linewidth]{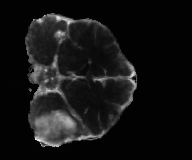}
\end{subfigure}%
\hfill
\begin{subfigure}[t]{.137\linewidth}  
\includegraphics[width=\linewidth]{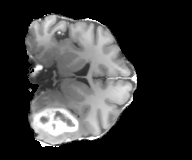}
\end{subfigure}%
%
%
%
%
%
\vspace{.4mm}
\begin{subfigure}[t]{.04\linewidth}  
\centering
\rotatebox{90}{\scriptsize T1}
\end{subfigure}%
\hfill
\begin{subfigure}[t]{.137\linewidth}  
\includegraphics[width=\linewidth]{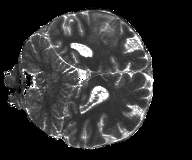}
\end{subfigure}%
\hfill
\begin{subfigure}[t]{.137\linewidth}  
\includegraphics[width=\linewidth]{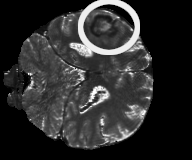}
\end{subfigure}%
\hfill
\begin{subfigure}[t]{.137\linewidth}  
\includegraphics[width=\linewidth]{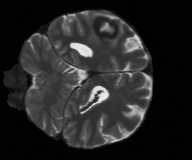}
\end{subfigure}%
\hfill
\begin{subfigure}[t]{.137\linewidth}  
\includegraphics[width=\linewidth]{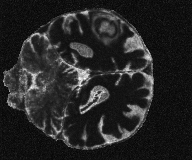}
\end{subfigure}%
\hfill
\begin{subfigure}[t]{.137\linewidth}  
\includegraphics[width=\linewidth]{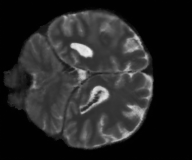}
\end{subfigure}%
\hfill
\begin{subfigure}[t]{.137\linewidth}  
\includegraphics[width=\linewidth]{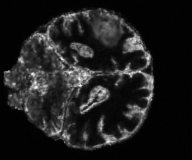}
\end{subfigure}%
\hfill
\begin{subfigure}[t]{.137\linewidth}  
\includegraphics[width=\linewidth]{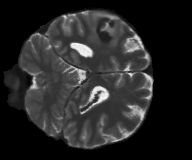}
\end{subfigure}%
\vspace{-.1mm}
\begin{subfigure}[t]{.04\linewidth}  
\centering
\rotatebox{90}{\scriptsize T1w}
\end{subfigure}%
\hfill
\begin{subfigure}[t]{.137\linewidth}  
\includegraphics[width=\linewidth]{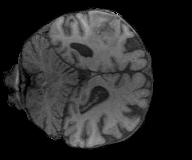}
\end{subfigure}%
\hfill
\begin{subfigure}[t]{.137\linewidth}  
\includegraphics[width=\linewidth]{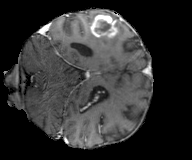}
\end{subfigure}%
\hfill
\begin{subfigure}[t]{.137\linewidth}  
\includegraphics[width=\linewidth]{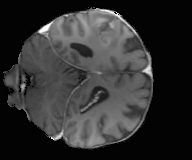}
\end{subfigure}%
\hfill
\begin{subfigure}[t]{.137\linewidth}  
\includegraphics[width=\linewidth]{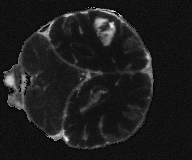}
\end{subfigure}%
\hfill
\begin{subfigure}[t]{.137\linewidth}  
\includegraphics[width=\linewidth]{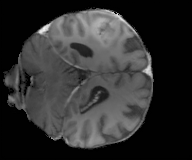}
\end{subfigure}%
\hfill
\begin{subfigure}[t]{.137\linewidth}  
\includegraphics[width=\linewidth]{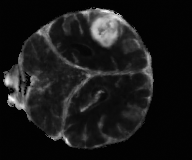}
\end{subfigure}%
\hfill
\begin{subfigure}[t]{.137\linewidth}  
\includegraphics[width=\linewidth]{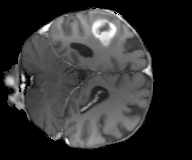}
\end{subfigure}%
%
%
%
%
%
\vspace{.4mm}
\begin{subfigure}[t]{.04\linewidth}  
\centering
\rotatebox{90}{\scriptsize T1}
\end{subfigure}%
\hfill
\begin{subfigure}[t]{.137\linewidth}  
\includegraphics[width=\linewidth]{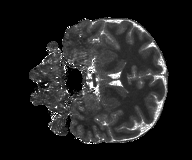}
\end{subfigure}%
\hfill
\begin{subfigure}[t]{.137\linewidth}  
\includegraphics[width=\linewidth]{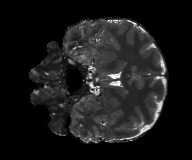}
\end{subfigure}%
\hfill
\begin{subfigure}[t]{.137\linewidth}  
\includegraphics[width=\linewidth]{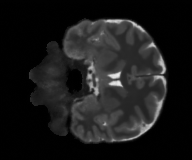}
\end{subfigure}%
\hfill
\begin{subfigure}[t]{.137\linewidth}  
\includegraphics[width=\linewidth]{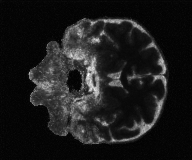}
\end{subfigure}%
\hfill
\begin{subfigure}[t]{.137\linewidth}  
\includegraphics[width=\linewidth]{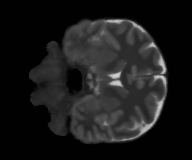}
\end{subfigure}%
\hfill
\begin{subfigure}[t]{.137\linewidth}  
\includegraphics[width=\linewidth]{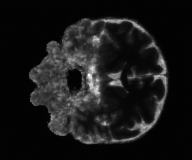}
\end{subfigure}%
\hfill
\begin{subfigure}[t]{.137\linewidth}  
\includegraphics[width=\linewidth]{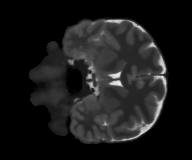}
\end{subfigure}%
\vspace{-.1mm}
\begin{subfigure}[t]{.04\linewidth}  
\centering
\rotatebox{90}{\scriptsize T1w}
\end{subfigure}%
\hfill
\begin{subfigure}[t]{.137\linewidth}  
\includegraphics[width=\linewidth]{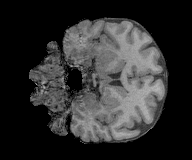}
\caption*{\scriptsize Pre}
\end{subfigure}%
\hfill
\begin{subfigure}[t]{.137\linewidth}  
\includegraphics[width=\linewidth]{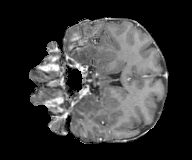}
\caption*{\scriptsize Post}
\end{subfigure}%
\hfill
\begin{subfigure}[t]{.137\linewidth}  
\includegraphics[width=\linewidth]{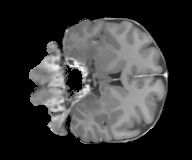}
\caption*{\scriptsize FM Mean}
\end{subfigure}%
\hfill
\begin{subfigure}[t]{.137\linewidth}  
\includegraphics[width=\linewidth]{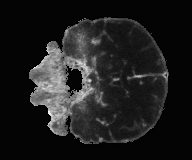}
\caption*{\scriptsize FM StdDev}
\end{subfigure}%
\hfill
\begin{subfigure}[t]{.137\linewidth}  
\includegraphics[width=\linewidth]{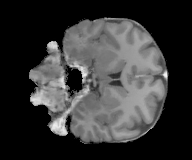}
\caption*{\scriptsize DM Mean}
\end{subfigure}%
\hfill
\begin{subfigure}[t]{.137\linewidth}  
\includegraphics[width=\linewidth]{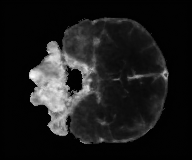}
\caption*{\scriptsize DM StdDev}
\end{subfigure}%
\hfill
\begin{subfigure}[t]{.137\linewidth}  
\includegraphics[width=\linewidth]{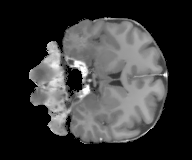}
\caption*{\scriptsize E2E}
\end{subfigure}%
\label{fig:final}
\caption{Visualization of FM, DM, and E2E prediction for T1 and T1w for  HGG and  MET test slices, and  HGG test slice without visible tumor (marked in white) (top to bottom).}
\label{fig:final}
\end{figure}
\section{Conclusion}
We have demonstrated encouraging results in Gadolinium enhancement prediction based on conditional generative networks, particularly for uncertainty quantification.
Finding a measure for comparing the prediction 
of T1 scans from qMRI with (usual) T1w MRI scans via segmentation, we have attested a promising T1 performance.
Still, these are preliminary findings based on a very limited dataset. 
Incorporating larger datasets and including either low-dose GBCA scans or additional information from quantitative T2 or diffusion data could be an exciting research direction.


\newpage
\section{Acknowledgements}
ES gratefully acknowledges funding by the Else Kröner-Fresenius-Stiftung.
MP and GS gratefully acknowledge the financial support by the German Research Foundation (DFG), GRK2260 BIOQIC project 289347353.

\section{Ethics Approval}
The study was approved by the ethics committee of the University Hospital Frankfurt (E8/20, E159/18). 

\bibliographystyle{abbrv}
\bibliography{references}
\end{document}